\documentclass[twocolumn]{aastex63}  
\usepackage{amsmath}

\shorttitle{Tidal Disruption of Sun-like Stars}
\shortauthors{Law-Smith et al.}

\begin{document}

\title{The Tidal Disruption of Sun-like Stars by Massive Black Holes}

\author[0000-0001-8825-4790]{Jamie Law-Smith}
\affiliation{Department of Astronomy and Astrophysics, University of California, Santa Cruz, CA 95064, USA}
\affiliation{Niels Bohr Institute, University of Copenhagen, Blegdamsvej 17, 2100 Copenhagen, Denmark}

\author[0000-0002-9809-8215]{James Guillochon}
\affiliation{Harvard-Smithsonian Center for Astrophysics, The Institute for Theory and Computation, 60 Garden Street, Cambridge, MA 02138, USA}

\author[0000-0003-2558-3102]{Enrico Ramirez-Ruiz}
\affiliation{Department of Astronomy and Astrophysics, University of California, Santa Cruz, CA 95064, USA}
\affiliation{Niels Bohr Institute, University of Copenhagen, Blegdamsvej 17, 2100 Copenhagen, Denmark}

\correspondingauthor{Jamie Law-Smith}
\email{lawsmith@ucsc.edu}

\begin{abstract}
We present the first simulations of the tidal disruption of stars with realistic structures and compositions by massive black holes (BHs). We build stars in the stellar evolution code MESA and simulate their disruption in the 3D adaptive-mesh hydrodynamics code FLASH, using an extended Helmholtz equation of state and tracking 49 elements. We study the disruption of a 1$M_\sun$ star and 3$M_\sun$ star at zero-age main sequence (ZAMS), middle-age, and terminal-age main sequence (TAMS). The maximum BH mass for tidal disruption increases by a factor of $\sim$2 from stellar radius changes due to MS evolution; this is equivalent to varying BH spin from 0 to 0.75. The shape of the mass fallback rate curves is different from the results for polytropes of \citet{2013ApJ...767...25G}. The peak timescale $t_{\rm{peak}}$ increases with stellar age, while the peak fallback rate $\dot{M}_{\rm{peak}}$ decreases with age, and these effects diminish with increasing impact parameter $\beta$. For a $\beta=1$ disruption of a 1$M_\sun$ star by a $10^6~M_\sun$ BH, from ZAMS to TAMS, $t_{\rm{peak}}$ increases from 30 to 54 days, while $\dot{M}_{\rm{peak}}$ decreases from 0.66 to $0.14~M_\sun$/yr. Compositional anomalies in nitrogen, helium, and carbon can occur before the peak timescale for disruptions of MS stars, which is in contrast to predictions from the ``frozen-in'' model. More massive stars can show stronger anomalies at earlier times, meaning that compositional constraints can be key in determining the mass of the disrupted star. The abundance anomalies predicted by these simulations provide a natural explanation for the spectral features and varying line strengths observed in tidal disruption events.
\end{abstract}


\section{Introduction}\label{sec:introduction}
The tidal disruption of a star by a massive black hole (BH) occurs when a star is knocked onto a nearly radial ``loss-cone'' orbit toward the BH by a chance encounter with another star. The flares resulting from the disruption can offer insight into otherwise quiescent massive BHs, the nuclear stellar populations that surround them, the physics of super-Eddington accretion, and the dynamical mechanisms operating in galactic centers. A detailed theoretical understanding of tidal disruptions is required to pry this information from observations. Pioneering theoretical work includes \citet{1975Natur.254..295H}, \citet{1983A&A...121...97C}, \citet{1988Natur.333..523R}, and \citet{1989ApJ...346L..13E}.

In this Letter, we present the first simulations of tidal disruptions of stars with realistic structures and compositions. We build stars using the 1D stellar evolution code MESA \citep{2011ApJS..192....3P} and calculate their disruption in the 3D adaptive-mesh hydrodynamics code FLASH \citep{2000ApJS..131..273F}. We track the elemental composition of the debris that falls back onto the black hole. We study the disruption of a $1 M_\sun$ star and $3 M_\sun$ star at three different ages.

A few dozen tidal disruption event (TDE) candidates have been observed thus far; see \citet{2015JHEAp...7..148K} and \citet{2017ApJ...838..149A} for a review of observations. Nearly all of their light curves (luminosity vs. time) are well fit by a simple scaling of mass fallback rate predictions from simulations \citep[e.g.,][]{2019ApJ...872..151M}, suggesting that circularization of the debris is prompt, and that the mass fallback rate has important discriminatory power in determining the key properties of an observed disruption \citep{2009ApJ...697L..77R}.

The shape of the mass fallback rate curve depends on the properties of the BH (mass, spin), the properties of the star (structure, mass), and the parameters of the disruption (impact parameter, orientation).  \citet{2013ApJ...767...25G} studied the impact of stellar structure and impact parameter on the mass fallback rate using $\gamma=4/3$ and $\gamma=5/3$ polytropic stellar structures. \citet{2019MNRAS.tmp.1458G} performed a parameter space study of relativistic tidal disruptions with spinning BHs for a $\gamma=5/3$ stellar structure. \citet{2019MNRAS.487..981G} recently simulated the disruption of a zero-age main sequence (ZAMS) star using moving-mesh hydrodynamics and studied the evolution of the stellar remnant, but did not track composition or study non-ZAMS stars.

Besides the shape of the light curve, spectroscopic information can provide clues as to the nature of the disrupted star. \citet{2016MNRAS.458..127K} predicted abundance anomalies in TDEs resulting from evolved stars. \citet{2018ApJ...857..109G} developed a simple framework, based on the work of \citet{2009MNRAS.392..332L} and \citet{2016MNRAS.458..127K}, to calculate the mass fallback rate for the disruption of stars of many masses and ages and to track the composition of the mass fallback. This is a useful framework that can be used to interpret spectroscopic observations of TDEs, but, as we discuss here, the simulations presented in this Letter make several different predictions from it.

An outstanding mystery in the field is that TDEs appear to occur preferentially in a rare type of galaxy \citep{2014ApJ...793...38A, 2016ApJ...818L..21F, 2017ApJ...850...22L, 2018ApJ...853...39G}. If we can determine the exact type of star that was disrupted in a TDE and build a demographic sample, we may be able to better understand this peculiar host galaxy preference. Separate from this, we may eventually be able to study the nuclear stellar populations in other galactic centers through tidal disruption.

TDEs can be used to obtain BH masses with comparable precision to the $M$--$\sigma$ relation \citep[e.g.,][]{2019ApJ...872..151M}. Simulations of tidal disruption using realistic stellar models will provide a better backbone for these fitting routines and a more accurate determination of all of the properties of the disruption. 

A diversity of stellar types can contribute to tidal disruptions from $10^5$--$10^9 M_\sun$ BHs; see the tidal disruption menu presented in \citet{2017ApJ...841..132L}. It is important to build a library of realistic tidal disruption simulations in order to extract the most information from the diversity of incoming and existing observations. The simulation framework we present in this Letter enables one to simulate the tidal disruption of any object that can be constructed in a stellar evolution code, allowing for the development of a library of tidal disruption simulations of stars with realistic structures and compositions.

This Letter is organized as follows. In Section~\ref{sec:methods} we discuss our methods. In Sections \ref{sec:structure} and \ref{sec:composition} we discuss our results with regard to stellar structure and composition respectively. In Section~\ref{sec:conclusion} we summarize and conclude.

\section{Methods}\label{sec:methods}

We build stars using the 1D stellar evolution code MESA and simulate their tidal disruption using FLASH, a 3D grid-based adaptive mesh refinement hydrodynamics code. For this study, we focus on the disruption of a $1 M_\sun$ star at ZAMS (0 Gyr), middle-age (4.8 Gyr), and TAMS (terminal-age main sequence; 8.4 Gyr), and a $3 M_\sun$ star at ZAMS (0 Gyr) and TAMS (0.3 Gyr). We simulate an encounter with a $10^6 M_\sun$ BH \citep[for non-relativistic encounters, other BH masses will simply scale the properties of the disruption; see e.g.][]{2013ApJ...767...25G} at a range of impact parameters from grazing encounters to full disruptions.

We use the following MESA setup\footnote{Inlists are available upon request.}: we start with a pre-MS model, use the \citet{2009ARA&A..47..481A} abundances (X=0.7154, Y=0.2703, and Z=0.0142), the \texttt{mesa\_49} nuclear network with the \texttt{jina} nuclear reaction rates preference \citep[from][]{2010ApJS..189..240C}, and \texttt{mixinglengthalpha=2.0} (this is the MESA default, and corresponds to setting the mixing length equal to twice the local pressure scale height\footnote{\citet{2016ApJ...817...54M} show that this is accurate for stellar masses up to 3$M_\sun$.}). We define TAMS as a central hydrogen fraction of $10^{-3}$. We track 49 elements, but in our results only show a few representative elements that have relatively high mass fractions. Full composition (and other) results will be made publicly available with the release of our tidal disruption library (in prep.).

\begin{figure*}[tbp!]
\epsscale{0.45}
\plotone{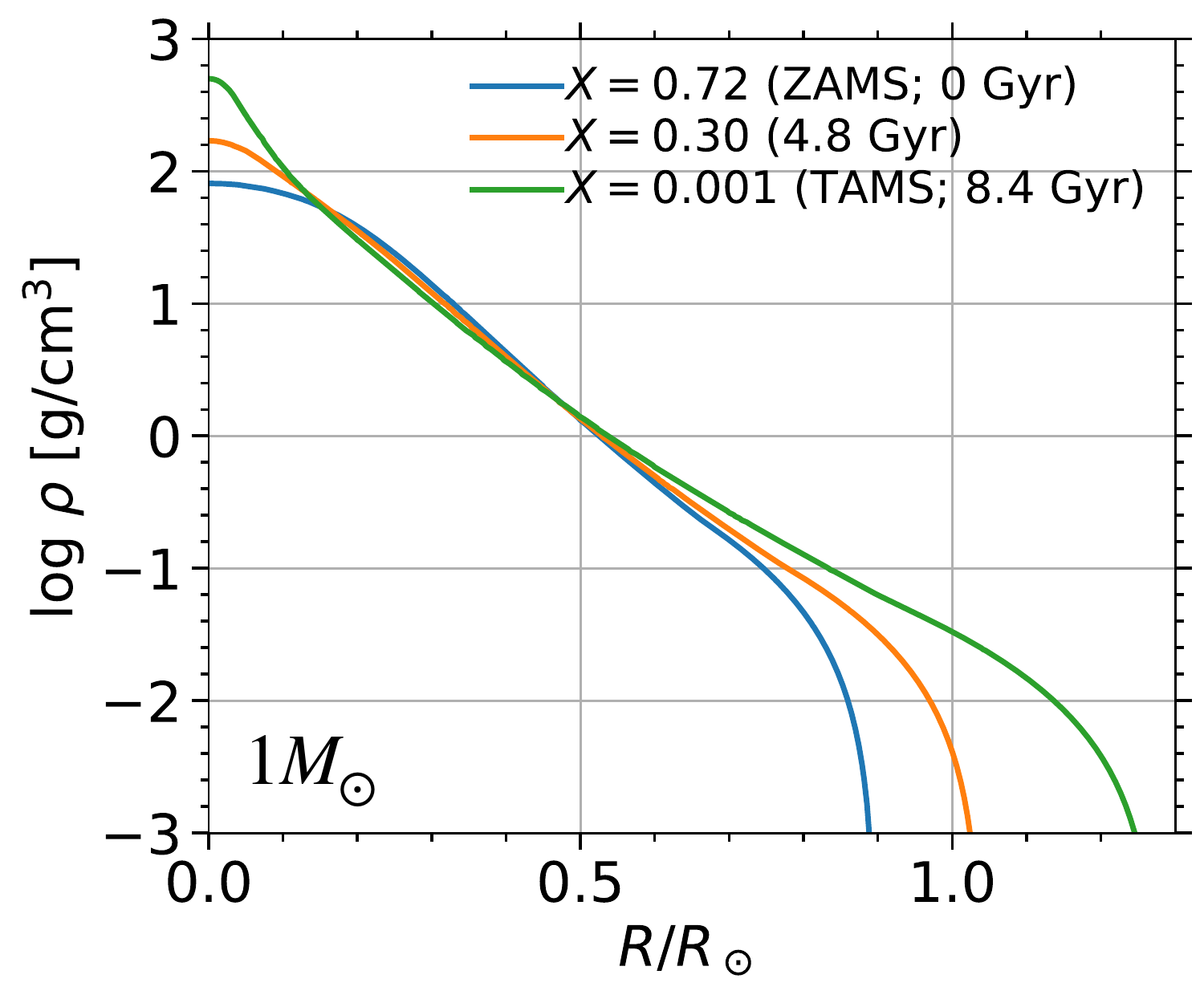}
\plotone{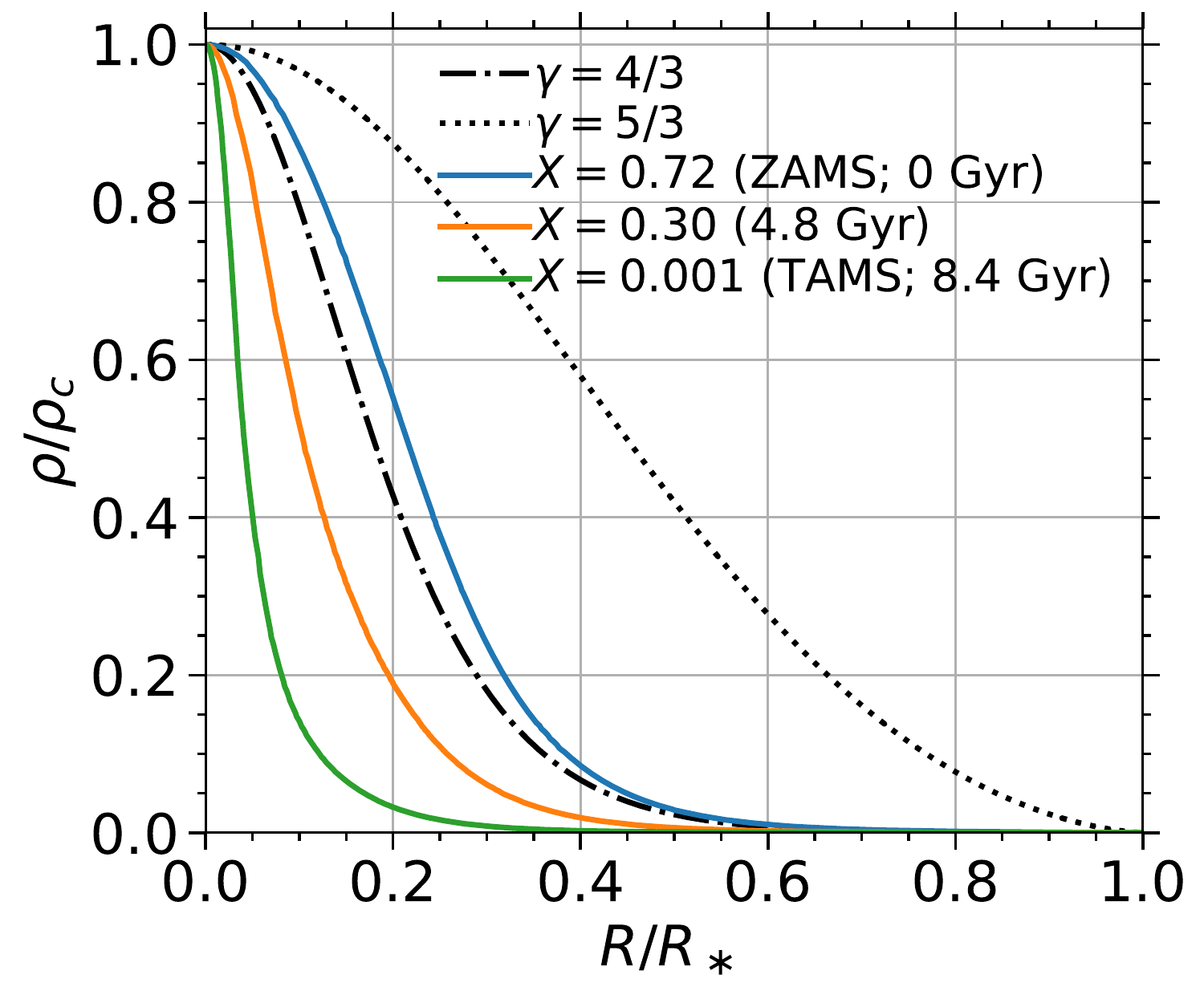}\\
\plotone{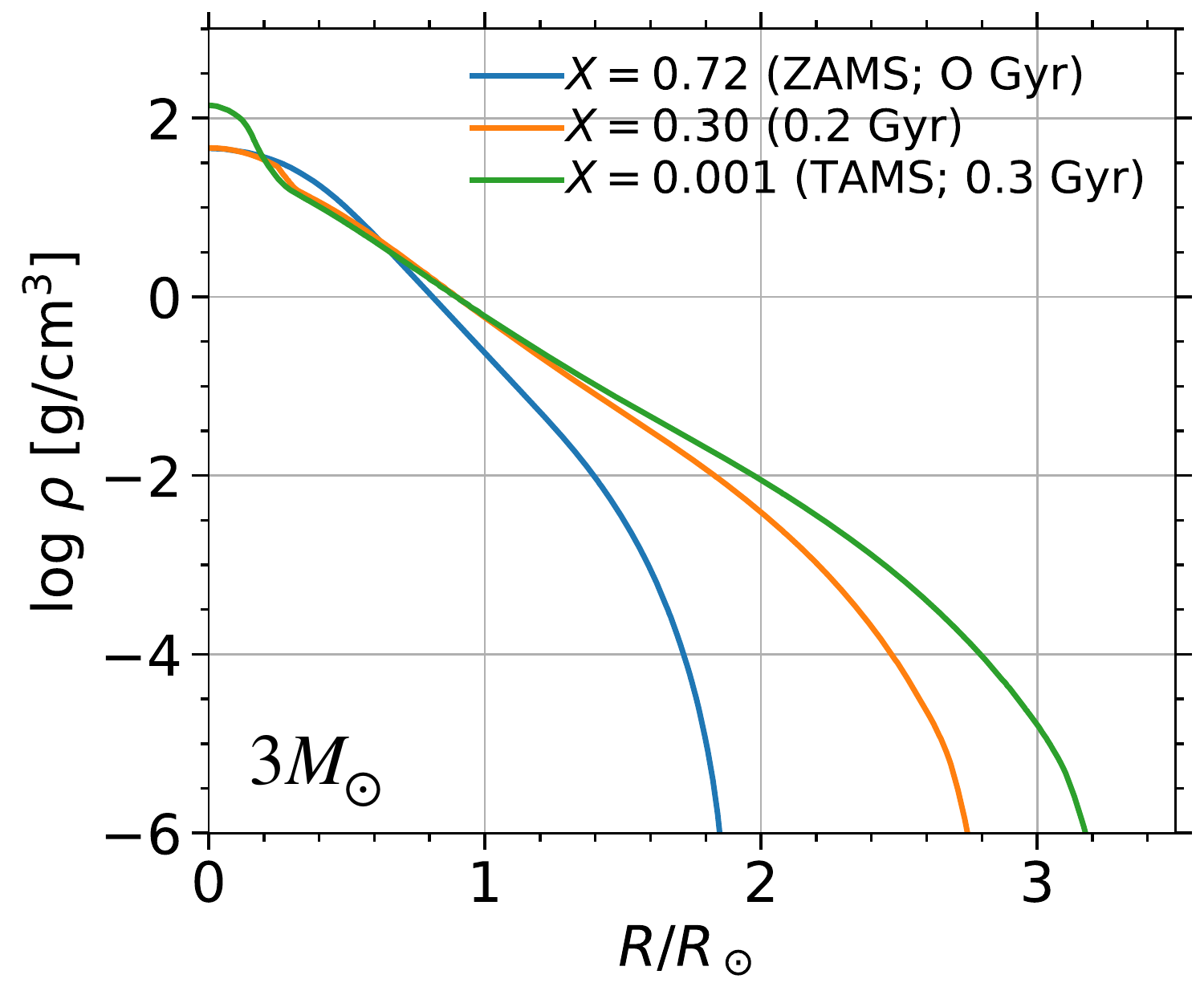}
\plotone{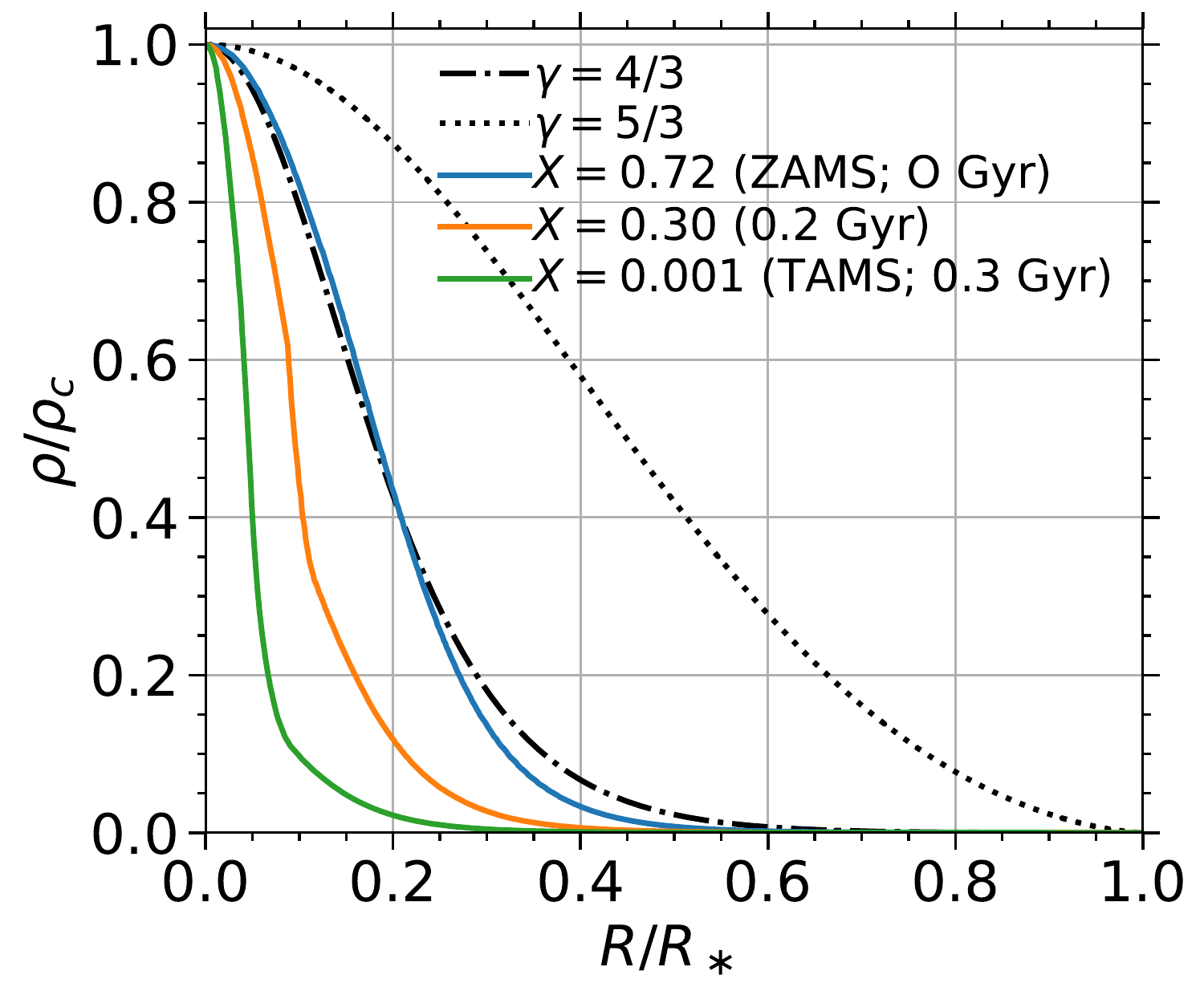}
\caption{
MESA density profiles for a $1 M_\sun$ star (top panels) and $3 M_\sun$ star (bottom panels) along their main sequence lifetimes. X is the central hydrogen mass fraction. Left panels: density vs. radius. Right panels: normalized to central density and stellar radius. Dashed and dotted lines show profiles for $\gamma=4/3$ and $\gamma=5/3$ polytropes respectively.
}
\label{fig:MESA}
\end{figure*}

We map the 1D profiles of density, pressure, temperature, and composition from MESA onto a 3D grid in FLASH, with initially uniform refinement. Some of the details of our FLASH setup are explained in \citet{2013ApJ...767...25G}. The important differences from this setup are that (1) we use an extended Helmholtz equation of state\footnote{This is an extension of the default FLASH Helmholtz table, based on \citet{2000ApJS..126..501T}, and is available at \href{http://cococubed.asu.edu/code_pages/eos.shtml}{http://cococubed.asu.edu/code\_pages/eos.shtml}.} rather than a polytropic equation of state, (2) we map a MESA profile onto the FLASH grid, and (3) we track the elemental composition of the debris for 49 elements. Our setup is Eulerian, centered on the rest frame of the star. Our domain is $1000 R_\star$ on a side, and we run our simulations until the stellar debris leaves the domain, typically $60$-$100~t_{\rm dyn}$ after the start of the simulation (the dynamical time of the star is defined as $t_{\rm dyn}=\sqrt{R_\star^3/GM_\star}$). 
This corresponds to 23-65 hours depending on the star and impact parameter. Note that the period of the most tightly bound debris in our simulations is (at shortest) $\approx 110$ hours, so no stream-stream collisions occur.
At initial maximum refinement, we have 131 cells across the initial diameter of the star. This is a factor of $\approx 2.6$ times better initial resolution than \citet{2013ApJ...767...25G}, which had $\approx 50$ cells across the initial diameter. The simulation retains this maximum refinement through pericenter and derefines as the debris spreads out. We refine based on density, relative to the maximum density in the simulation. All cells within $10^{-5}$ of the maximum density have the same refinement (are maximally refined). The simulations presented in this Letter have a maximum total number of blocks of $4.8\times10^4$. There are $8^3=512$ cells per block, so this translates to $2.5\times10^7$ maximum cells in the simulation.

\begin{figure*}[tbp!]
\epsscale{0.37}
\plotone{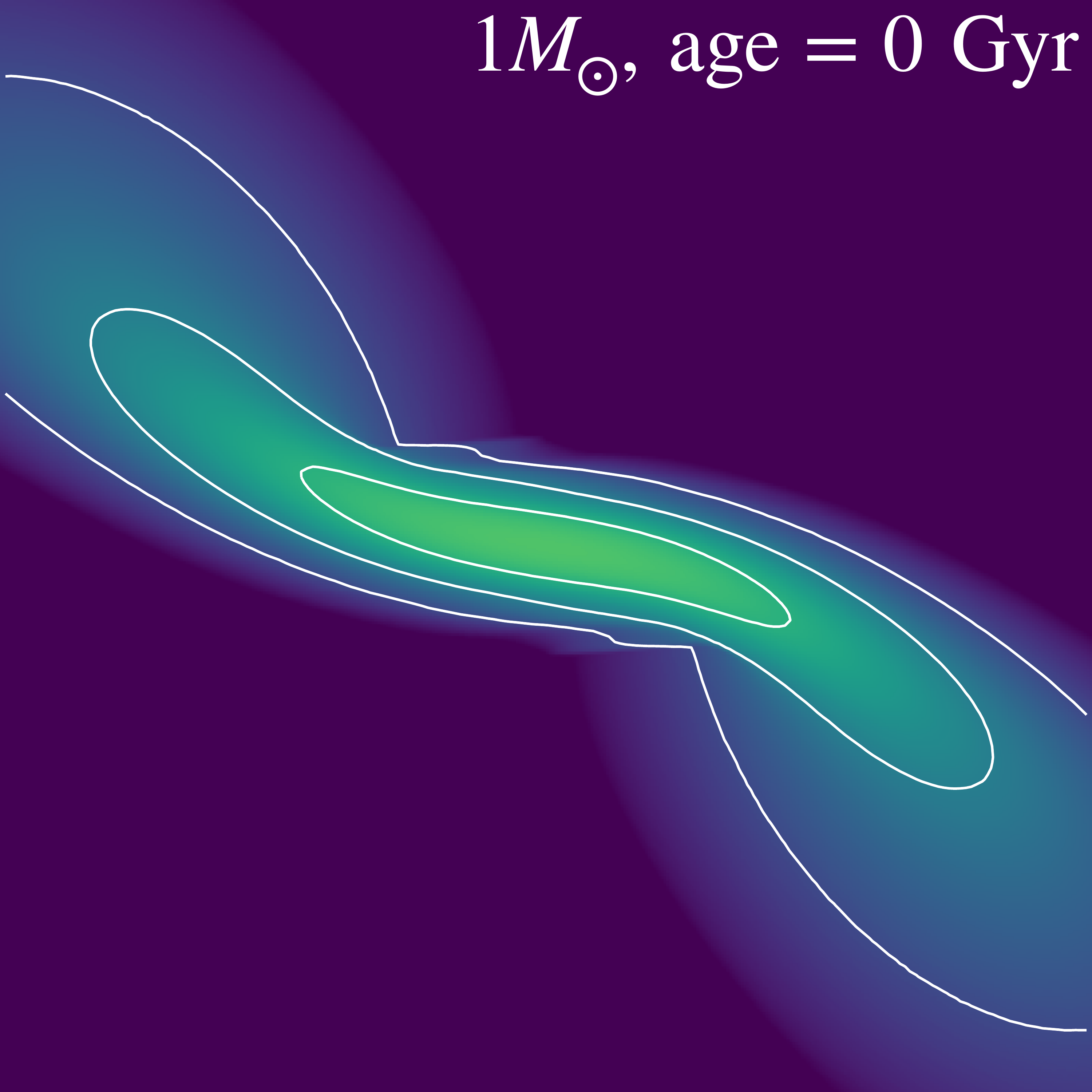}
\plotone{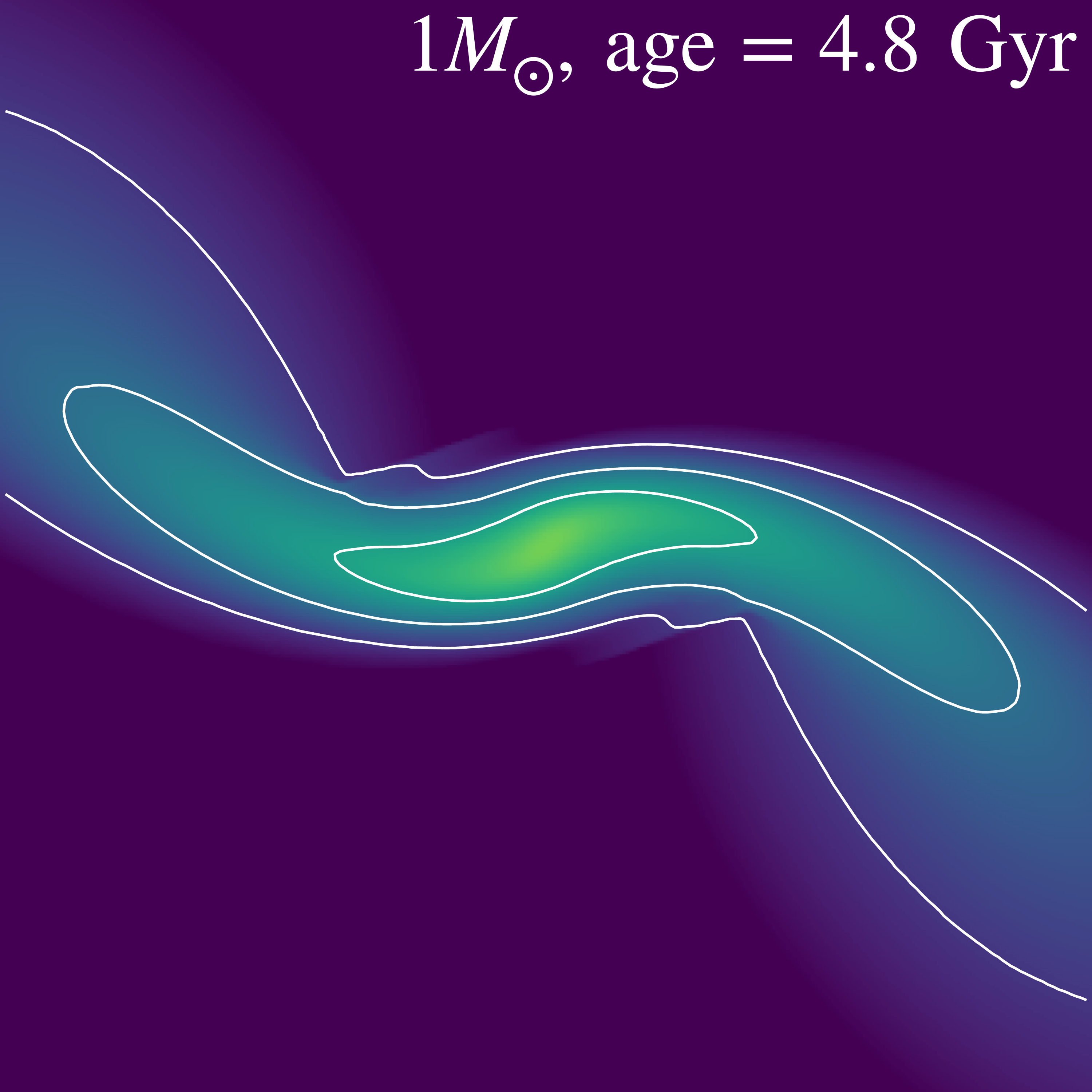}
\plotone{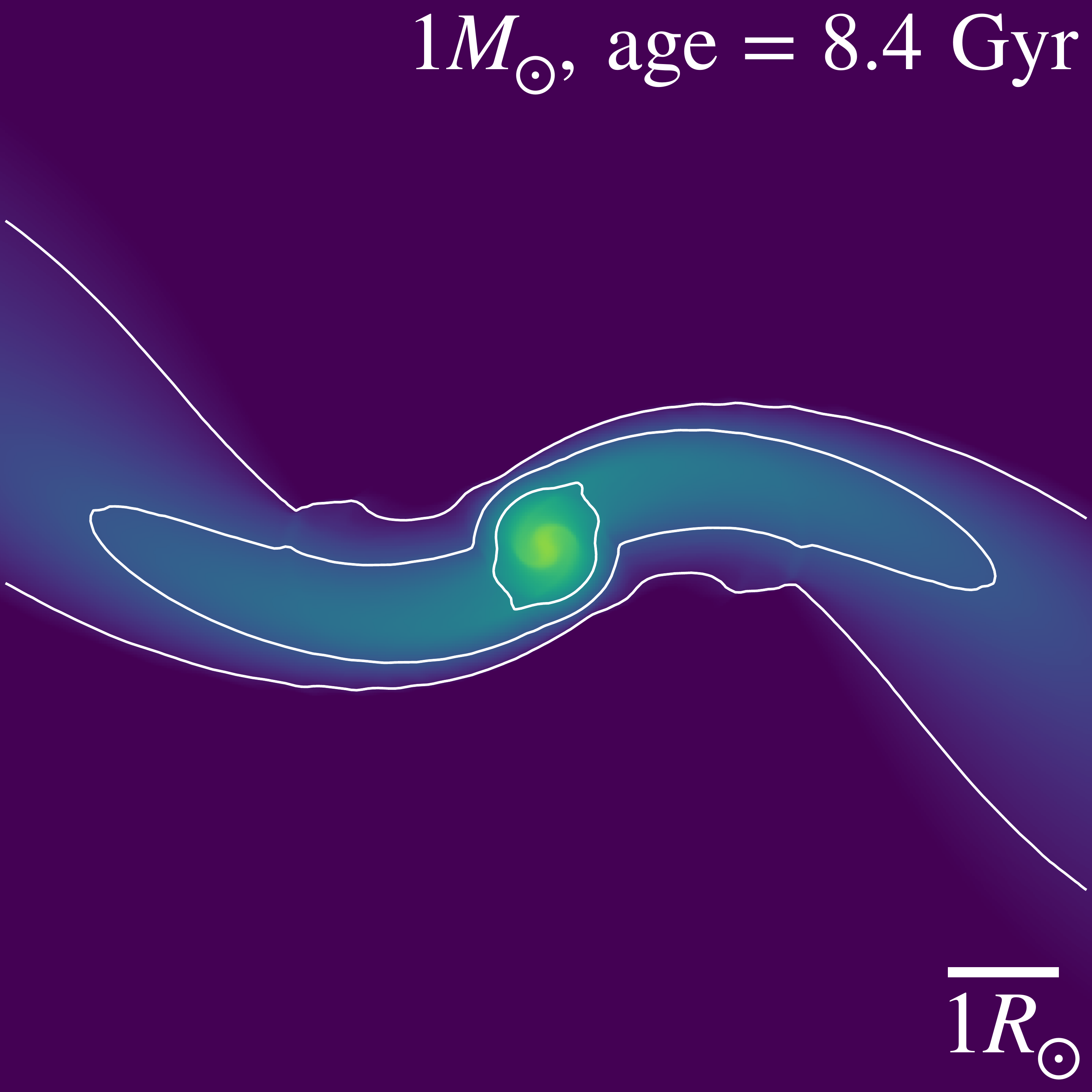}
\caption{
2D slices in the orbital plane of a $\beta=2$ encounter with a $10^6 M_\sun$ BH for a $1 M_\sun$ star at ZAMS, middle-age, and TAMS, at $\approx 3~t_{\rm dyn}$ after pericenter. Color corresponds to density and contours are equally spaced in the logarithm of the density (at $\rho=1,10^{-1},10^{-2}\ {\rm g/cm^3}$). Videos of the simulations are available at \href{https://www.youtube.com/channel/UCShahcfGrj5dOZTTrOEqSOA}{this URL}.
}
\label{fig:snapshots}
\end{figure*}

The impact parameter $\beta \equiv r_{\rm t}/r_{\rm p}$ is defined as the ratio of the tidal radius,
\begin{equation}
r_{\rm t} \equiv (M_{\rm BH}/M_\star)^{1/3} R_\star,
\end{equation}
to the pericenter distance, $r_{\rm p}$. 
Note that the tidal radius is defined using the stellar radius (not necessarily 1$R_\sun$ for a 1$M_\sun$ star), so that the same impact parameter for different stellar ages corresponds to different pericenter distances.
The most relativistic encounter shown in this work is a $\beta=3$ disruption of a ZAMS Sun; here $r_{\rm p} \simeq 14 GM_{\rm BH}/c^2$. In this regime, relativistic effects on the rate of return of the fallback material are minor \citep{2017MNRAS.469.4483T,2019GReGr..51...30S}.
Note also that in our simulations the tidal radius is $100 R_\star$, meaning that the BH enters the computational domain as it moves through pericenter. This does not lead to any issues vis-a-vis capture by the event horizon, as the star's deformation through pericenter only extends to a few $R_\star$, and further, the pericenter passage takes place on the star's dynamical timescale. Put more precisely, the minimum angular momentum of the tidal debris is much greater than the threshold for capture.
We begin the simulations at $r=10~r_{\rm t}$, where tidal effects are negligible.\footnote{Cf., for example, \citet{2019MNRAS.487..981G}, whose simulations start at $5~r_{\rm t}$.} We then relax the star onto the grid for $5~t_{\rm dyn}$ before beginning the parabolic BH orbit evolution. We verify that the stellar profiles after this relaxation process are very similar to the intial input MESA profiles \citep[see also e.g.][]{2017ApJ...841..132L}.

We calculate the mass fallback rate ($\dot M$) to the BH by first calculating the spread in binding energy $dM/dE$ of each cell in our simulation. We smooth the $dM/dE$ distribution with a Gaussian filter, as it is noisy due to our fine binning, then convert this distribution to an $\dot M$ curve through Kepler's third law. Our $\dot M$ curves are derived at the last time at which all of the stellar debris is within the domain, $40$-$80~t_{\rm dyn}$ after pericenter; Figure~10 of \citet{2013ApJ...767...25G}, which shows $\dot M$ curves up to $550~t_{\rm dyn}$ after pericenter, demonstrates that our $\dot M$ curves are accurate for the timescales we are interested in for this work. We tested that our setup can reproduce the \citet{2013ApJ...767...25G} $\dot M$ and $\Delta M$ results for polytropes at a few different impact parameters. We verified the resolution convergence of our results by running a subset of our simulations with twice or four times the maximum number of blocks stated above, finding no appreciable difference.

\section{Stellar Structure}\label{sec:structure}

In this section we consider the structure evolution of a $1 M_\sun$ star and $3 M_\sun$ star along their main sequence lifetimes as representative examples. Stars with $M \gtrsim 3 M_\sun$ will be very rare as TDEs due to their short main sequence lifetimes; stars with $M \lesssim 0.6 M_\sun$, on the other hand, will not significantly evolve over the age of the universe.

Figure~\ref{fig:MESA} shows density profiles from MESA for a $1 M_\sun$ star and $3 M_\sun$ star along their main sequence lifetimes. From ZAMS to TAMS, the Sun's central density increases by a factor of $\approx 6$, from 80 g/cm$^3$ to 500 g/cm$^3$, and its radius increases by a factor of $\approx 1.4$, from $0.9 R_\sun$ to $1.3 R_\sun$. A $3 M_\sun$ star's radius increases by a factor of 1.75 over its MS lifetime. Normalized to central density and stellar radius, the profile of a $\gamma=4/3$ polytrope is in rough agreement with that of a ZAMS Sun and in better agreement with that of a ZAMS $3 M_\sun$ star, though is not a good match for non-ZAMS stars.

The density profile of a star determines its susceptibility to tidal disruption. Figure~\ref{fig:snapshots} shows 2D slices in the orbital plane from simulations of the disruption of the Sun at three different ages (ZAMS, middle-age, and TAMS) at the same impact parameter ($\beta=2$). For the ZAMS Sun this is a full disruption, whereas for the TAMS Sun this is a grazing encounter in which a core survives.

\begin{figure*}[tbp]
\epsscale{0.45}
\plotone{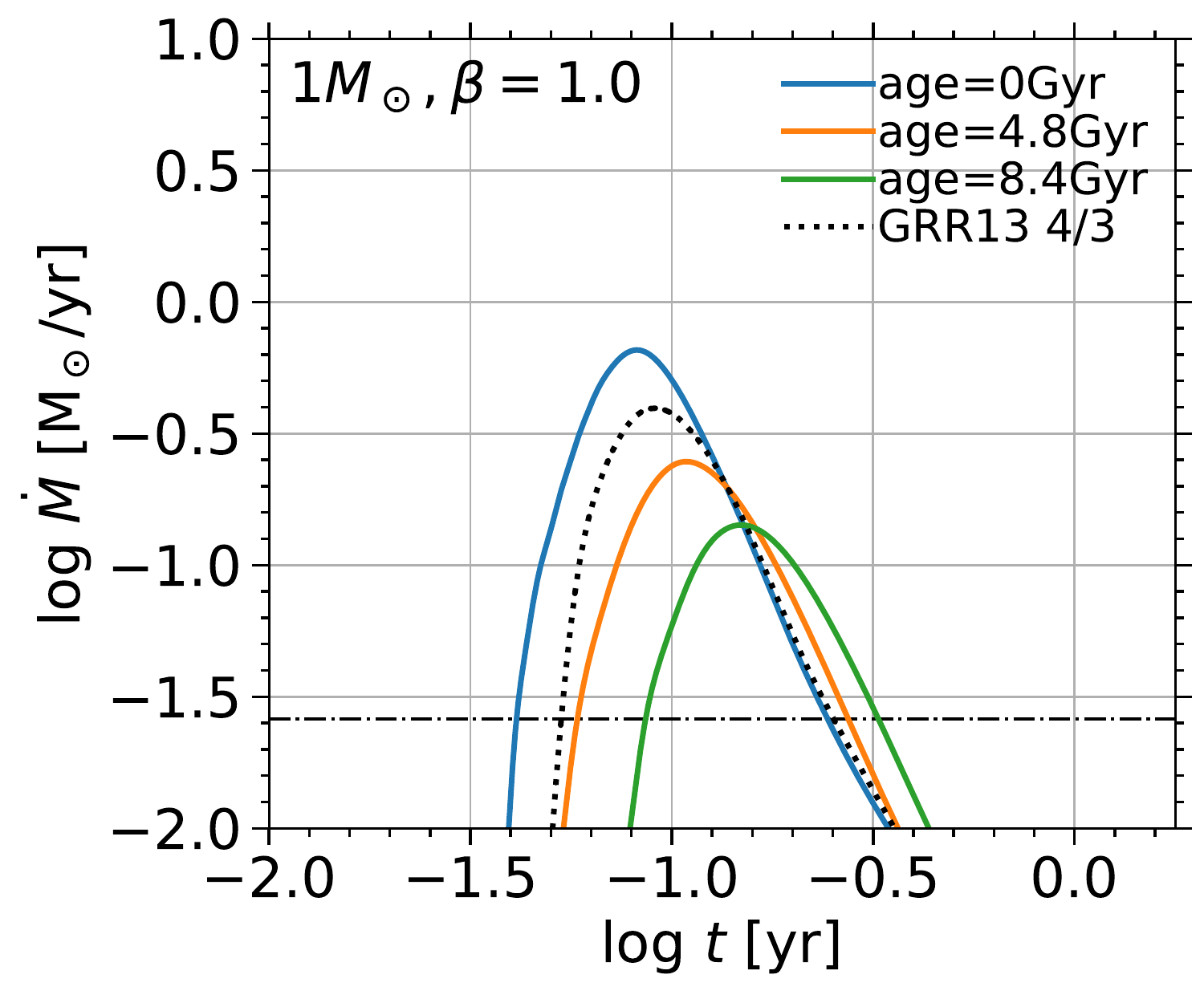}{(a)}
\plotone{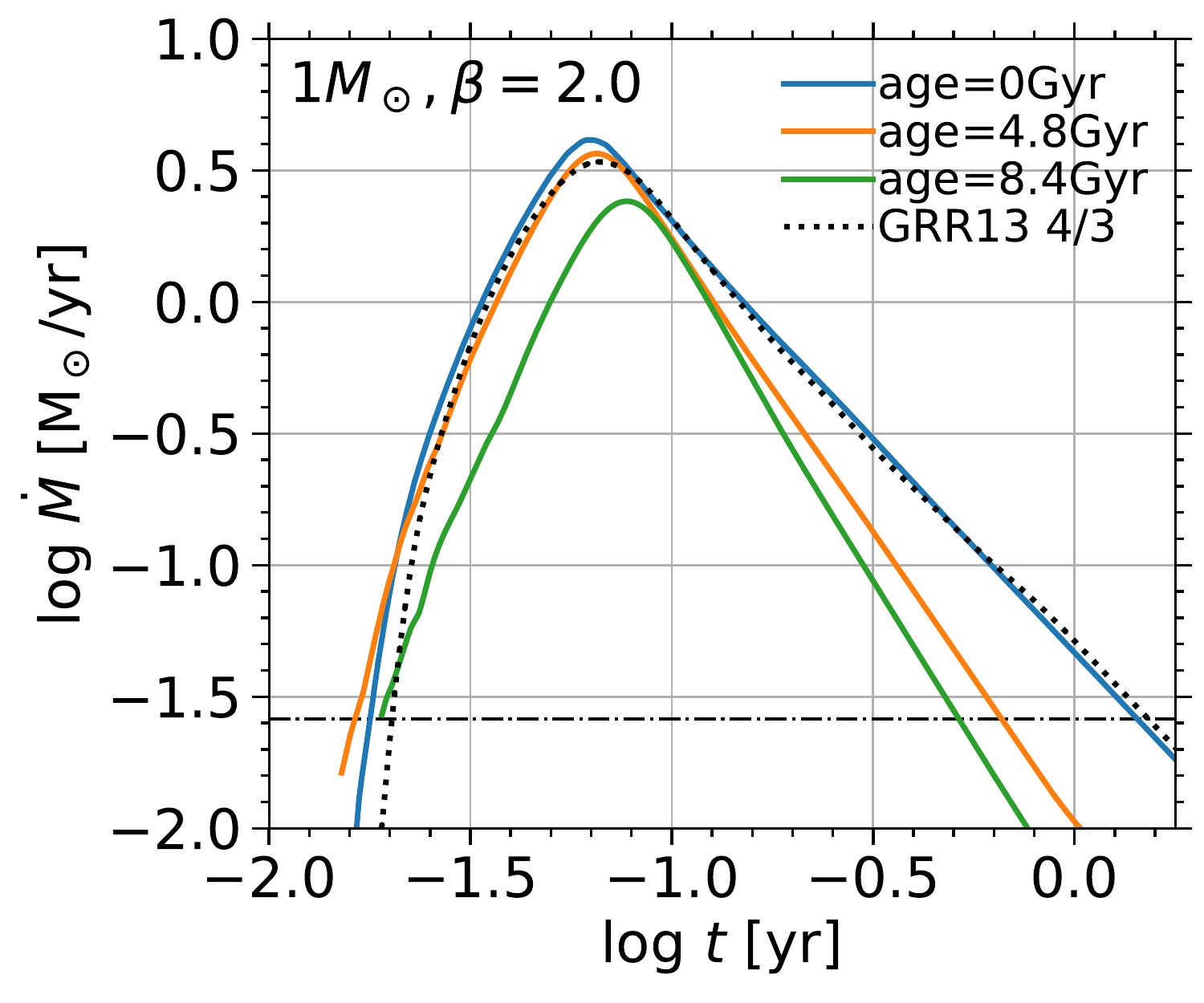}{(b)}
\plotone{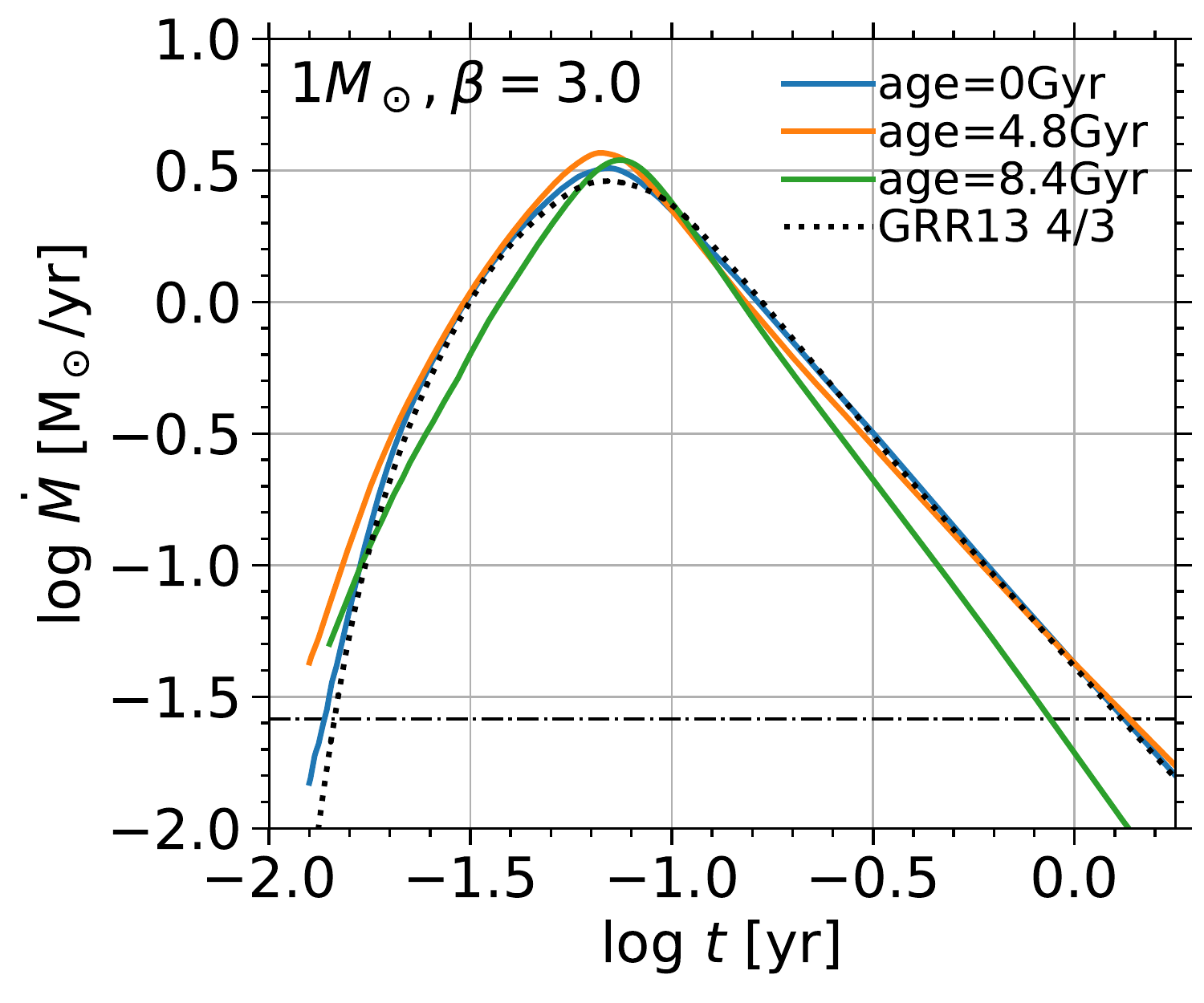}{(c)}
\plotone{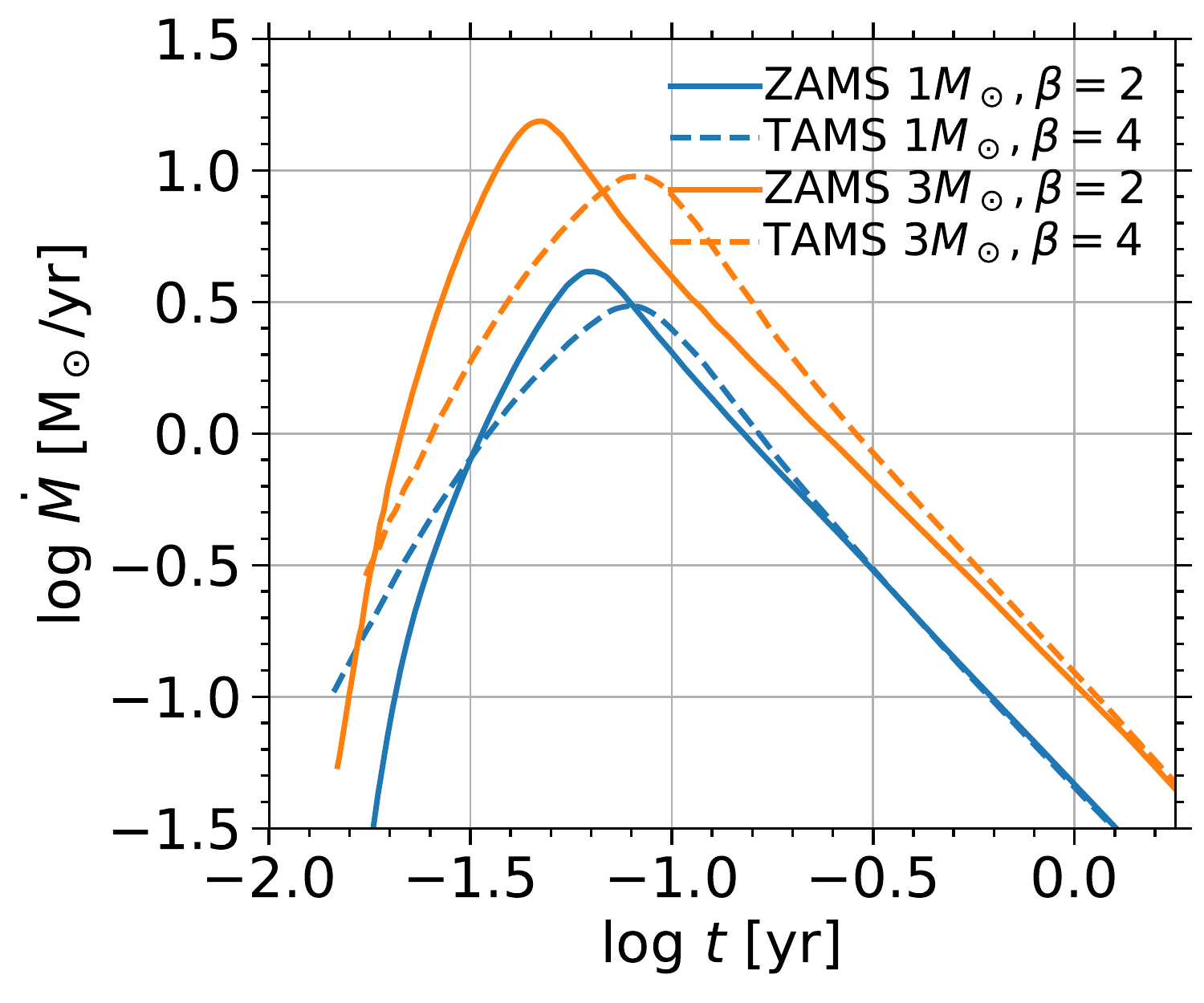}{(d)}
\caption{
Panels (a), (b), (c): mass fallback rate to the BH as a function of time for the disruption of a $1 M_\sun$ star at three different ages and impact parameters by a $10^6 M_\sun$ BH. Panels are grouped by impact parameter $\beta$. The result for a $\gamma=4/3$ polytrope from \citet{2013ApJ...767...25G}, scaled to the radius of the ZAMS Sun, is in dotted black. The Eddington limit for this BH, assuming a radiative efficiency of $\epsilon=0.1$ and an electron scattering opacity of $\kappa=0.34$ cm$^2$ g$^{-1}$, is shown by the dot-dashed line. Panel (d): mass fallback rate for full disruptions of a $1 M_\sun$ star and $3 M_\sun$ star at ZAMS and TAMS.
\label{fig:mdots}
}
\end{figure*}

As the density profile of a star changes, so does the mass fallback rate to the BH resulting from its disruption. Panels (a), (b), and (c) of Figure~\ref{fig:mdots} show the mass fallback rate $\dot M$ to the BH as a function of time for the disruption of the Sun for three impact parameters at three different ages (results here for the $3 M_\sun$ star show similar trends). Panels are grouped by impact parameter.

Older stars are more centrally concentrated and thus more difficult to fully disrupt, resulting in higher critical impact parameters for full disruption. At a fixed $\beta$, the amount of mass lost $\Delta M$ decreases with stellar age.\footnote{Note however that at {\it fixed pericenter distance} $r_{\rm p}$, because older stars have larger radii, for low-$\beta$ partial disruptions the mass lost is larger for older stars.} The shape of the $\dot M$ curve also changes: at a fixed $\beta$ (for the $\beta$'s shown in this work), the slope of the $\dot M$ curve after peak becomes steeper with stellar age---this is mostly easily seen for the $\beta=2$ disruptions. This behavior was also observed for partial disruptions of a given polytrope in \citet{2013ApJ...767...25G}.

The time of peak of the mass fallback rate, $t_{\rm peak}$, increases with stellar age (i.e., younger stars can provide faster flares) and this effect diminishes with increasing $\beta$. The peak mass fallback rate, $\dot M_{\rm peak}$, decreases with stellar age and this effect diminishes at high $\beta$. For $\beta=1$, from ZAMS to TAMS for the Sun, $t_{\rm peak}$ increases from 30 days to 54 days, while $\dot M_{\rm peak}$ decreases from $0.66~M_\sun$/yr to $0.14~M_\sun$/yr. For $\beta=2$, $t_{\rm peak}$ increases from 23 to 28 days, while $\dot M_{\rm peak}$ decreases from $4.1~M_\sun$/yr to $2.4~M_\sun$/yr. For $\beta=3$, the peak properties for the three ages are more similar. Fitting formulae will be provided with a more extensive parameter study in impact parameter, mass, and age in future work.

We compare to the simulation results of \citet{2013ApJ...767...25G} for a $\gamma=4/3$ polytrope (the $\gamma=5/3$ simulations are more dissimilar), scaled to the radius of the ZAMS Sun. For $\beta=1$, the $\gamma=4/3$ simulation is in rough agreement but does not match any of the ages particularly well. For $\beta=2$, the $\gamma=4/3$ simulation more closely matches the ZAMS Sun, but does not capture the shape of the $\dot M$ curve for the middle-age or TAMS Suns. For $\beta=3$, the $\gamma=4/3$ simulation is a better approximation of the general shape for all three ages, but is a worse match for the TAMS Sun.

\begin{figure*}[tbp]
\epsscale{1}
\plotone{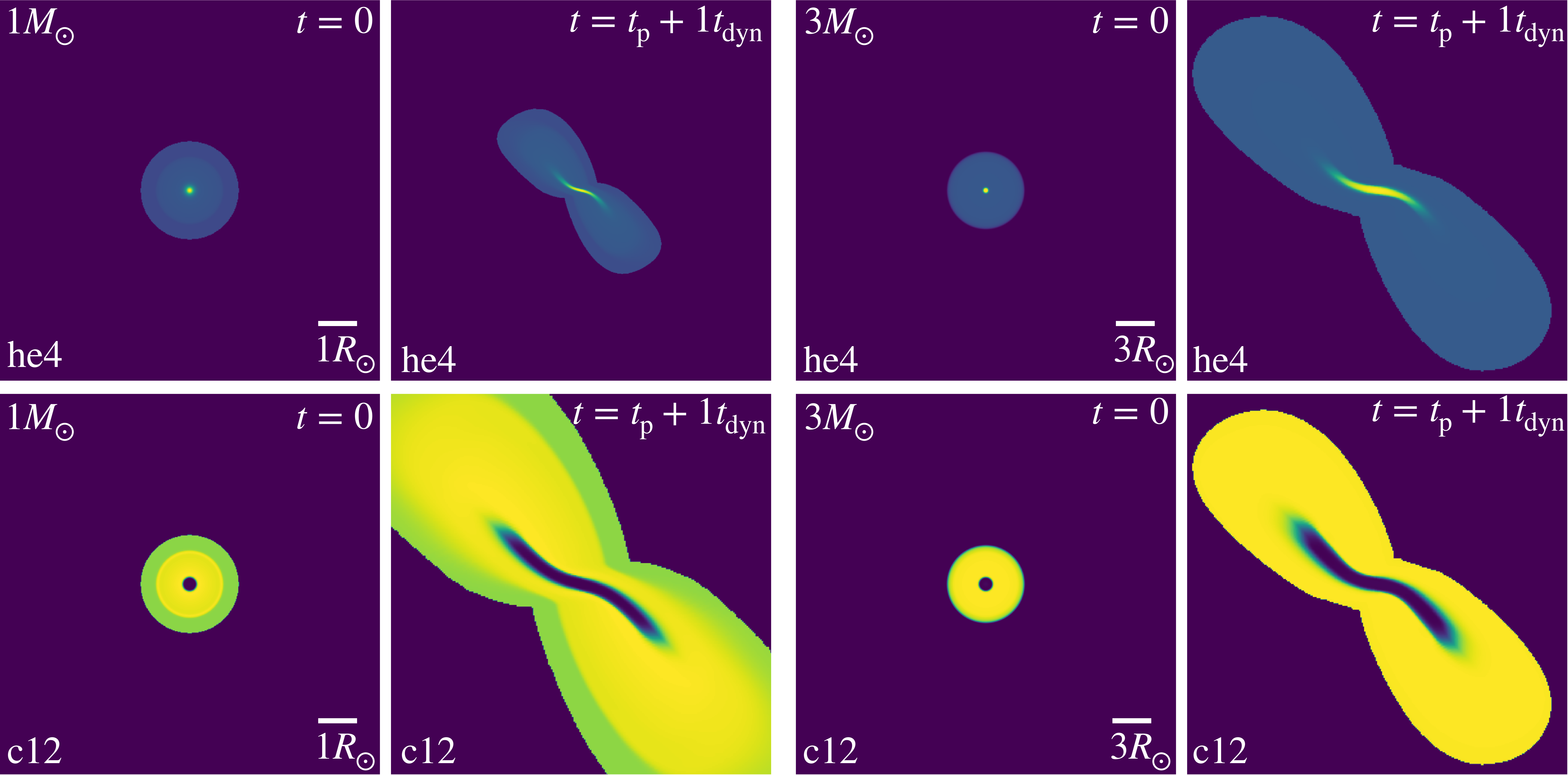}
\caption{
2D slices in the orbital plane of the mass fractions of helium and carbon for a $\beta=4$ disruption of a TAMS $1 M_\sun$ and $3 M_\sun$ star, at the start of the simulation and at $\approx 1~t_{\rm dyn}$ after pericenter. Color corresponds to the mass fraction of the element, with yellow being higher. The panels are normalized separately. The right panels in each group of four have a density cut of $10^{-4}$ g/cm$^3$.
\label{fig:rings}
}
\end{figure*}

The shape of the $\dot M$ curve is useful in determining the properties of the disruption when fitting to observed events \citep[e.g.,][]{2019ApJ...872..151M}, and a full library of tidal disruption simulations using realistic stellar profiles will improve these determinations. However, there are certain difficulties and degeneracies which can be resolved by incorporating more information. For example, for full disruptions, there is not a large variation in $t_{\rm peak}$ with the mass of the star. In panel (d) of Figure~\ref{fig:mdots}, we compare the mass fallback rate for full disruptions\footnote{We conducted a preliminary parameter-space study to determine the approximate impact parameters for full disruption (these are $\beta \approx 2$ for ZAMS 1$M_\sun$, $\beta \approx 3$ for middle-age 1$M_\sun$, $\beta \approx 4$ for TAMS 1$M_\sun$, $\beta \approx 2$ for ZAMS 3$M_\sun$, and $\beta \approx 4$ for TAMS 3$M_\sun$, with approximate uncertainty $\pm 0.5$).} of a $1 M_\sun$ star and $3 M_\sun$ star at ZAMS and TAMS. At a given evolutionary state, the normalization of the $\dot M$ curve changes with mass but the peak timescale does not vary much: it decreases by $\approx 5$ days from a ZAMS $1 M_\sun$ to $3 M_\sun$ star. Age can increase the spread: from a ZAMS to TAMS $3 M_\sun$ star, $t_{\rm peak}$ increases by $\approx 12$ days. This implies that determinations of BH masses are expected to be relatively robust, as the uncertainties associated with stellar mass and age do not greatly alter the shape of the resultant $\dot M$ curves. On the other hand, using light curves alone might be insufficient to effectively identify the nature of the disrupted star.
Using compositional information as a second axis can significantly improve our determinations of the properties of the disruption and it is to this issue that we now turn our attention.

\section{Composition}\label{sec:composition}

Tracking compositional information in our hydrodynamical simulations captures the mixing of previously sequestered regions within a star. This mixing affects the timing and composition of the debris returning to the BH. Figure~\ref{fig:rings} shows 2D slices of the mass fractions of helium and carbon for a $\beta=4$ disruption of a TAMS $1 M_\sun$ and $3 M_\sun$ star. Both the helium enhancement and the depletion of carbon in the stars' cores are mixed into the tidal tails. 
Note that while nuclear burning occurs primarily via the pp chain in the 1$M_\sun$ star and the CNO cycle in the 3$M_\sun$ star, carbon is similarly depleted in the cores of both of the stars; this is primarily because carbon is depleted during pre-MS evolution for the 3$M_\sun$ star.

\begin{figure*}[tbp]
\epsscale{0.381}
\plotone{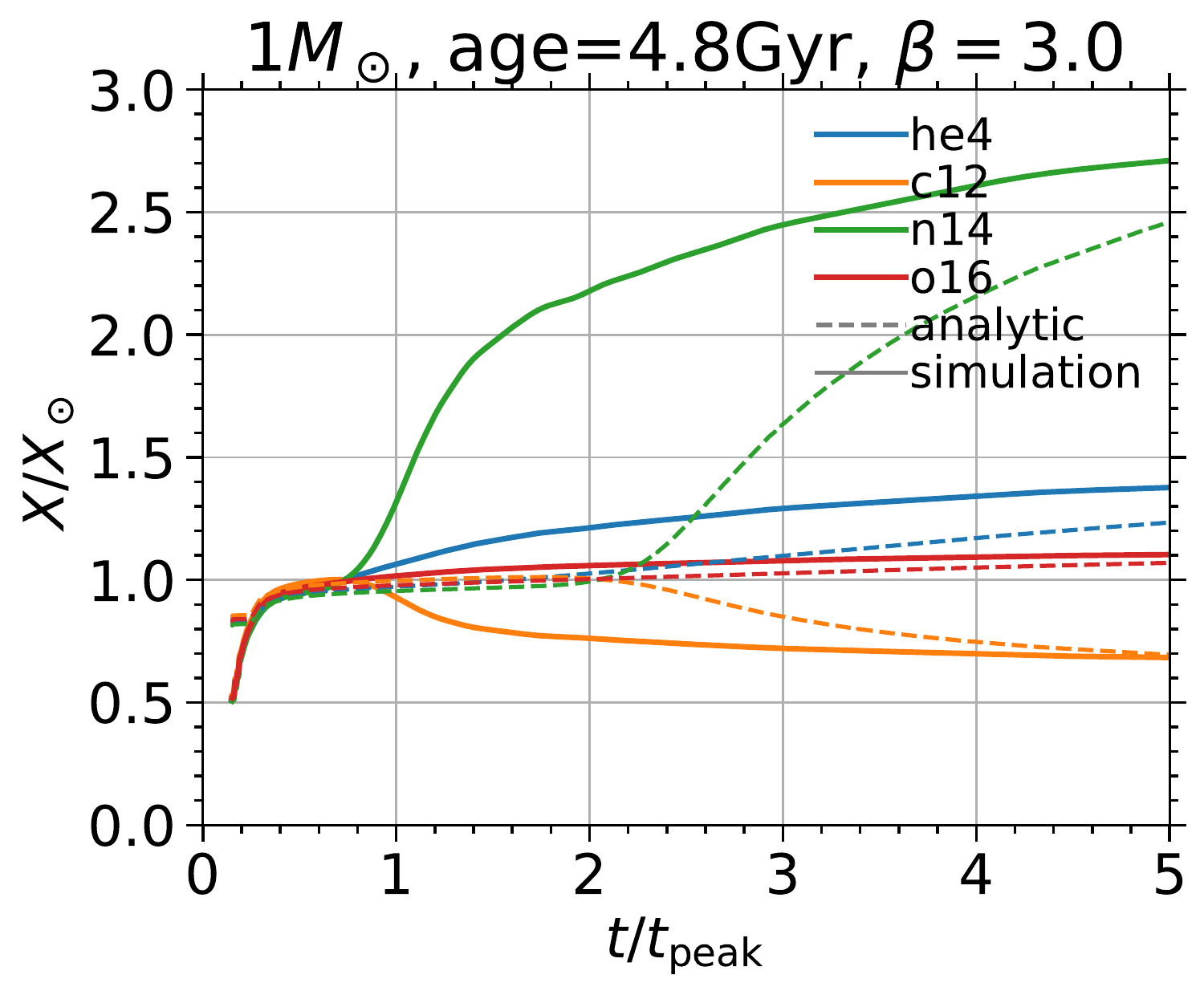}
\plotone{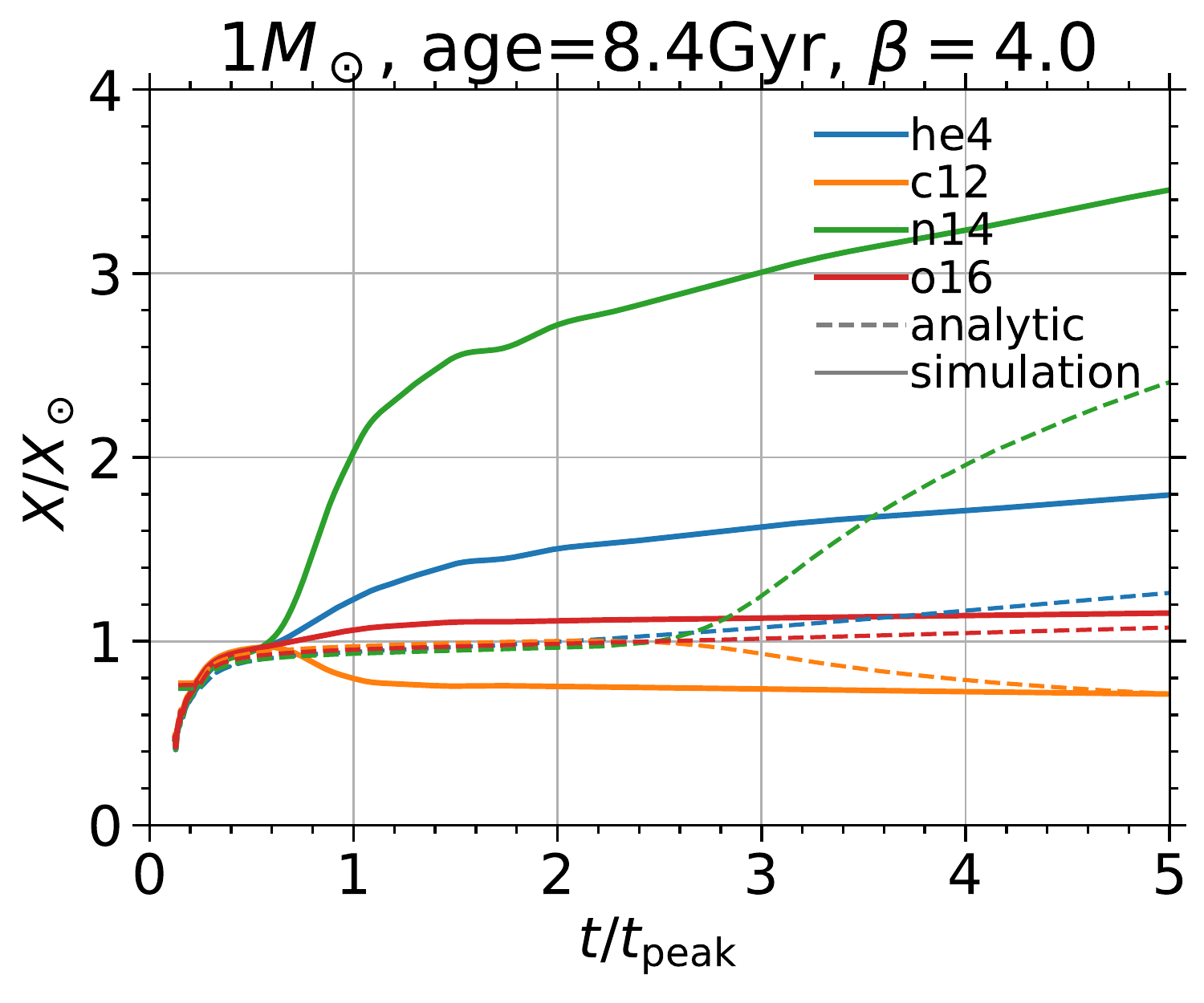}
\plotone{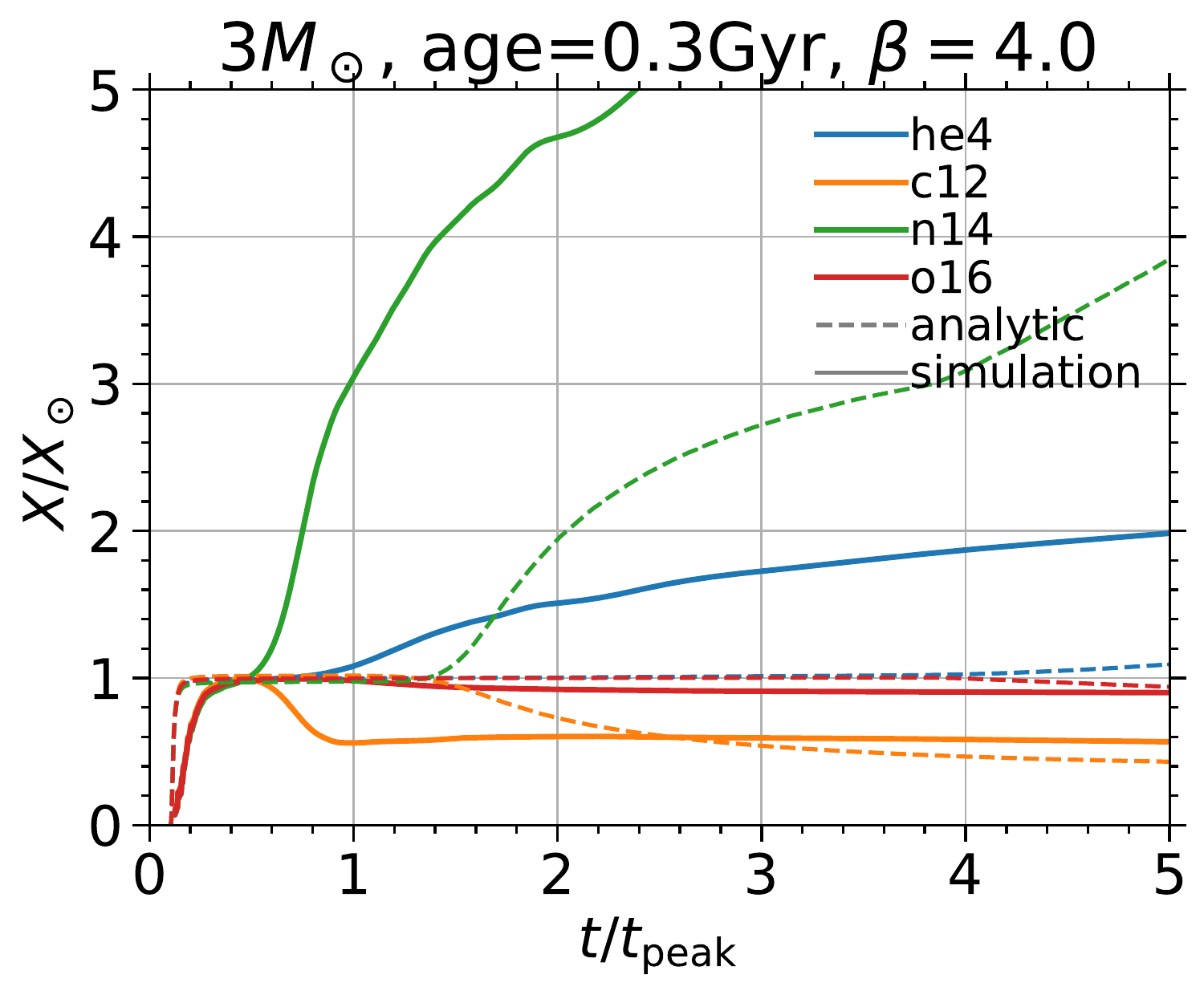}
\caption{
Composition (relative to solar) of the fallback material to pericenter as a function of time (relative to the peak of the mass fallback rate). The panels from left to right show full disruptions of a middle-age Sun, a TAMS Sun, and a TAMS $3M_\sun$ star. Solid lines are hydrodynamic simulation results and dashed lines are analytic results from \citet{2018ApJ...857..109G}.
\label{fig:composition}
}
\end{figure*}

Figure~\ref{fig:composition} shows the composition of stellar material returning to pericenter as a function of time for three full disruptions: a $\beta=3$ disruption of a middle-age Sun, a $\beta=4$ disruption of a TAMS Sun, and a $\beta=4$ disruption of a TAMS $3M_\sun$ star. We define
\begin{equation}
    \frac{X}{X_\sun} = \frac{\dot M_X / \dot M_{\rm H}}{M_X/M_{\rm H, \sun}},
\end{equation}
where $X$ is a given element, H refers to hydrogen, and the denominator is the abundance of $X$ relative to hydrogen in the Sun. Refer to Figure~3 of \citet{2018ApJ...857..109G} for the compositional evolution of the Sun along its main sequence lifetime.

For a ZAMS Sun, for all impact parameters, $X/X_\sun \simeq 1$ for all elements as a function of time. This follows from the fact that a 1$M_\sun$ star is nearly perfectly homogeneous at ZAMS. This is not, however, true of a ZAMS 3$M_\sun$ star (see below). For stars that have evolved along the MS, $X/X_\sun$ can be $\approx 1$ for low-$\beta$ (grazing) encounters that only strip the outside layers of the star unaffected by nuclear burning. Deeper encounters of non-ZAMS stars show non-solar fallback abundances. In general, abundance variations manifest as an increase in nitrogen and helium and a decrease in carbon over time, with an increase or decrease in oxygen depending on the mass of the star. Other elements, such as neon, sodium, and magnesium, show an increase over time. The relative strength and timing of these anomalies is a function of the mass and age of the star and the impact parameter of the disruption. More massive stars, older stars, and deeper encounters result in stronger abundance anomalies at earlier times.

For full disruptions of a middle-age Sun, TAMS Sun, and TAMS $3M_\sun$ star, abundance anomalies appear before the time of peak fallback rate. Helium, carbon, nitrogen, and oxygen (among many other elements with lower mass fractions) can all be enhanced or depleted before $t_{\rm peak}$. Abundance anomalies can also appear before peak for partial disruptions---for example, this occurs for a $\beta=3$ disruption of a TAMS Sun. These early variations are particularly encouraging for observations of the signatures of these kinds of disruptions. Additionally, the N/C ratio shows even stronger variations than the above individual elements. Though not shown here, nitrogen, helium, and oxygen abundances continue to rise/decrease for more than one year after peak (for the disruption of a $1 M_\sun$ star by a $10^6 M_\sun$ BH, until $\approx 6$ years after peak). That is, at late times, the elemental abundances asymptote to fixed values \citep[this late-time behavior was also predicted in our analytic framework,][]{2018ApJ...857..109G}.

Focusing on the time at which nitrogen is enhanced by a factor of 3 as a diagnostic of the timing of abundance anomalies: this occurs at $\approx 10~t_{\rm peak}$ for a middle-age Sun, at $\approx 3~t_{\rm peak}$ for a TAMS Sun, and at $\approx t_{\rm peak}$ for a TAMS $3M_\sun$ star. There is a similar trend in other elements---for example, for a TAMS $3M_\sun$ star, carbon is depleted by a factor of $\approx 2$ at $t_{\rm peak}$. Though not shown here, the full disruption of a ZAMS $3M_\sun$ star exhibits abundance variations in nitrogen and carbon, but at a lower level than for the TAMS star. A TAMS $3M_\sun$ star shows stronger abundance variations at earlier times compared to a $1 M_\sun$ star; thus, abundance anomalies increase with age and $M_\star$ at a fixed $t/t_{\rm peak}$. If TDEs occur (on average) for stars of the same age in a given nuclear stellar cluster, then more massive stars will provide stronger abundance anomalies.
Another determinant of mass is oxygen: oxygen is enhanced for the $1 M_\sun$ star but depleted for the $3 M_\sun$ star.

If strong abundance variations are observed at early times in a TDE (in the simple picture that abundance variations beget spectral features), this is a sign of a higher-$\beta$ disruption of a higher-mass star. Note, however, that the prospect of identifying the nature of the disrupted star is further complicated by $\beta$. For example, if more modest abundance variations are observed, it may be difficult to discern between a low-$\beta$ disruption of a higher-mass star and a high-$\beta$ disruption of a lower-mass star. A full library of simulations with fitting formulae will help break this degeneracy.

We also compare our results to predictions from the analytic framework of \citet{2018ApJ...857..109G}. The simulations show stronger abundance variations at early times. We note that over longer timescales ($t \gtrsim 10~t_{\rm peak}$), the analytic framework is in good general agreement with the simulations, but we focus on timescales near peak here, as these are the most relevant to current observations. The analytic framework is useful for predicting broad features of the composition of the fallback material for many stellar masses and ages, but is limited in that it cannot probe the $\beta$ parameter space (as it is only applicable to full disruptions) and more importantly, it does not capture the deformation and spin-up of the star at pericenter (it assumes that the star arrives intact to pericenter, at which point the binding energy is ``frozen-in'').  
The fact that at pericenter, the star is typically spun-up to a large fraction of its breakup angular velocity and has a highly distorted shape, as well as the subsequent mixing of debris as the disruption evolves, account for the differences between the analytic model and the simulations. See \citet{2019MNRAS.485L.146S} for a more detailed examination of the differences between the ``frozen-in'' model and hydrodynamical simulations.

\section{Conclusion}\label{sec:conclusion}

We built stars with realistic stellar profiles and elemental compositions in MESA and simulated their tidal disruption in FLASH, using a Helmholtz equation of state and tracking the composition of the debris. The shape of the mass fallback rate curves and the tidal susceptibility for a star at different ages along its main sequence lifetime differ from results for polytropes from \citet{2013ApJ...767...25G}. $t_{\rm peak}$ increases with stellar age, while $\dot M_{\rm peak}$ decreases with age, and these effects diminish with increasing impact parameter. Significant mixing and rotation of the debris occurs during disruption, leading to abundance anomalies appearing before the peak of the mass fallback rate for some disruptions. In the fallback debris for non-ZAMS stars, nitrogen and helium are enhanced and carbon is depleted relative to solar. Abundance variations are stronger at earlier times for older and more massive stars.

Strong nitrogen and a lack of carbon (C III) features, and in two cases strong oxygen features, have been observed in the four TDEs with UV spectra extending to these wavelengths: ASASSN-14li \citep{2016ApJ...818L..32C}, iPTF16fnl \citep{2018MNRAS.473.1130B}, iPTF15af \citep{2019ApJ...873...92B}, and AT 2018dyb \citep{2019arXiv190303120L}. These features are naturally explained by our simulations as the tidal disruptions of non-ZAMS stars. A stronger N/C ratio at an earlier time relative to peak (such as the nitrogen feature observed at $t \approx 1.2~t_{\rm peak}$ in iPTF16fnl) indicates that a flare arose from the disruption of a more massive star. Time-resolved spectroscopy extending into the UV will be very useful for fitting to simulations and determining the mass of the disrupted star.

It is important to note that stellar evolution along the MS leads to significant changes in the density profile of the star, but also in its radius. The Sun's radius changes from $0.9 R_\sun$ to $1.3 R_\sun$ from ZAMS to TAMS. The maximum black hole mass for disruption (assuming the same $\beta$) increases by a factor of 1.75. A $3 M_\sun$ star's radius changes from 1.9 to 3.3 $R_\sun$ from ZAMS to TAMS; the maximum BH mass increases by a factor of 2.3.
So the uncertainty on maximum BH mass from stellar evolution is $\sim 2$. From Figure~1 of \citet{2012PhRvD..85b4037K}, a factor of $\sim 2$ in maximum BH mass is equivalent to a change in black hole spin of 0 to 0.75 (from a spin of 0.75 to 1, the maximum BH mass changes by a factor of 4). The uncertainty from stellar evolution can therefore be of the same order as the uncertainty from BH spin---this is important as it is BH spin that determines the cutoff of the TDE rate as a function of BH mass in Figure~4 of \citet{2012PhRvD..85b4037K} \citep[this is also Figure~4 of][]{2019GReGr..51...30S}.

We plan to construct a library of tidal disruption simulations of stars built in MESA, for different stellar masses and ages, tracking composition information. As the present study shows, these simulations can reveal important behavior not captured by earlier models. Now that the sample of TDEs with high quality observations has grown to a few dozen (and continues to grow), it is very important to construct a library of tidal disruption simulations of realistic stars with fitting formulae for important disruption quantities. In using simulations such as these to fit light curve and spectral information, it may be possible to accurately determine the mass of the disrupted star, as well as provide more accurate fits for all of the other properties of the disruption (BH mass, spin, efficiency, etc.).

Additionally, the framework developed in this Letter can be used to study the surviving remnants of tidal disruption. These objects can have unique compositions and internal dynamics. For example, the late-time checkpoint of the surviving star could be used as an input to MESA for future stellar evolution calculations.

\acknowledgements
We thank Ryan Foley, Nathaniel Roth, Jane Dai, and Brenna Mockler for useful conversations, as well as the anonymous referee for insightful comments. We thank the Niels Bohr Institute for its hospitality while part of this work was completed, and acknowledge the Kavli Foundation and the DNRF for supporting the 2017 Kavli Summer Program. J.L.-S. and E.R.-R. acknowledge support from NASA ATP grant NNX14AH37G, NSF grant AST-1615881, the Heising-Simons Foundation and the Danish National Research Foundation (DNRF132).

\bibliography{export-bibtex}

\begin{thebibliography}{}
\expandafter\ifx\csname natexlab\endcsname\relax\def\natexlab#1{#1}\fi
\providecommand{\url}[1]{\href{#1}{#1}}
\providecommand{\dodoi}[1]{doi:~\href{http://doi.org/#1}{\nolinkurl{#1}}}
\providecommand{\doeprint}[1]{\href{http://ascl.net/#1}{\nolinkurl{http://ascl.net/#1}}}
\providecommand{\doarXiv}[1]{\href{https://arxiv.org/abs/#1}{\nolinkurl{https://arxiv.org/abs/#1}}}

\bibitem[{{Arcavi} {et~al.}(2014){Arcavi}, {Gal-Yam}, {Sullivan}, {Pan},
  {Cenko}, {Horesh}, {Ofek}, {De Cia}, {Yan}, \& {Yang}}]{2014ApJ...793...38A}
{Arcavi}, I., {Gal-Yam}, A., {Sullivan}, M., {et~al.} 2014, \apj, 793, 38,
  \dodoi{10.1088/0004-637X/793/1/38}

\bibitem[{{Asplund} {et~al.}(2009){Asplund}, {Grevesse}, {Sauval}, \&
  {Scott}}]{2009ARA&A..47..481A}
{Asplund}, M., {Grevesse}, N., {Sauval}, A.~J., \& {Scott}, P. 2009, \araa, 47,
  481, \dodoi{10.1146/annurev.astro.46.060407.145222}

\bibitem[{{Auchettl} {et~al.}(2017){Auchettl}, {Guillochon}, \&
  {Ramirez-Ruiz}}]{2017ApJ...838..149A}
{Auchettl}, K., {Guillochon}, J., \& {Ramirez-Ruiz}, E. 2017, \apj, 838, 149,
  \dodoi{10.3847/1538-4357/aa633b}

\bibitem[{{Blagorodnova} {et~al.}(2019){Blagorodnova}, {Cenko}, {Kulkarni},
  {Arcavi}, {Bloom}, {Duggan}, {Filippenko}, {Fremling}, {Horesh}, \&
  {Hosseinzadeh}}]{2019ApJ...873...92B}
{Blagorodnova}, N., {Cenko}, S.~B., {Kulkarni}, S.~R., {et~al.} 2019, \apj,
  873, 92, \dodoi{10.3847/1538-4357/ab04b0}

\bibitem[{{Brown} {et~al.}(2018){Brown}, {Kochanek}, {Holoien}, {Stanek},
  {Auchettl}, {Shappee}, {Prieto}, {Morrell}, {Falco}, \&
  {Strader}}]{2018MNRAS.473.1130B}
{Brown}, J.~S., {Kochanek}, C.~S., {Holoien}, T.~W.~S., {et~al.} 2018, \mnras,
  473, 1130, \dodoi{10.1093/mnras/stx2372}

\bibitem[{{Carter} \& {Luminet}(1983)}]{1983A&A...121...97C}
{Carter}, B., \& {Luminet}, J.~P. 1983, \aap, 121, 97

\bibitem[{{Cenko} {et~al.}(2016){Cenko}, {Cucchiara}, {Roth}, {Veilleux},
  {Prochaska}, {Yan}, {Guillochon}, {Maksym}, {Arcavi}, \&
  {Butler}}]{2016ApJ...818L..32C}
{Cenko}, S.~B., {Cucchiara}, A., {Roth}, N., {et~al.} 2016, \apj, 818, L32,
  \dodoi{10.3847/2041-8205/818/2/L32}

\bibitem[{{Cyburt} {et~al.}(2010){Cyburt}, {Amthor}, {Ferguson}, {Meisel},
  {Smith}, {Warren}, {Heger}, {Hoffman}, {Rauscher}, {Sakharuk}, {Schatz},
  {Thielemann}, \& {Wiescher}}]{2010ApJS..189..240C}
{Cyburt}, R.~H., {Amthor}, A.~M., {Ferguson}, R., {et~al.} 2010, \apjs, 189,
  240, \dodoi{10.1088/0067-0049/189/1/240}

\bibitem[{{Evans} \& {Kochanek}(1989)}]{1989ApJ...346L..13E}
{Evans}, C.~R., \& {Kochanek}, C.~S. 1989, \apj, 346, L13,
  \dodoi{10.1086/185567}

\bibitem[{{French} {et~al.}(2016){French}, {Arcavi}, \&
  {Zabludoff}}]{2016ApJ...818L..21F}
{French}, K.~D., {Arcavi}, I., \& {Zabludoff}, A. 2016, \apj, 818, L21,
  \dodoi{10.3847/2041-8205/818/1/L21}

\bibitem[{{Fryxell} {et~al.}(2000){Fryxell}, {Olson}, {Ricker}, {Timmes},
  {Zingale}, {Lamb}, {MacNeice}, {Rosner}, {Truran}, \&
  {Tufo}}]{2000ApJS..131..273F}
{Fryxell}, B., {Olson}, K., {Ricker}, P., {et~al.} 2000, \apjs, 131, 273,
  \dodoi{10.1086/317361}

\bibitem[{{Gafton} \& {Rosswog}(2019)}]{2019MNRAS.tmp.1458G}
{Gafton}, E., \& {Rosswog}, S. 2019, \mnras, 1458,
  \dodoi{10.1093/mnras/stz1530}

\bibitem[{{Gallegos-Garcia} {et~al.}(2018){Gallegos-Garcia}, {Law-Smith}, \&
  {Ramirez-Ruiz}}]{2018ApJ...857..109G}
{Gallegos-Garcia}, M., {Law-Smith}, J., \& {Ramirez-Ruiz}, E. 2018, \apj, 857,
  109, \dodoi{10.3847/1538-4357/aab5b8}

\bibitem[{{Goicovic} {et~al.}(2019){Goicovic}, {Springel}, {Ohlmann}, \&
  {Pakmor}}]{2019MNRAS.487..981G}
{Goicovic}, F.~G., {Springel}, V., {Ohlmann}, S.~T., \& {Pakmor}, R. 2019,
  \mnras, 487, 981, \dodoi{10.1093/mnras/stz1368}

\bibitem[{{Graur} {et~al.}(2018){Graur}, {French}, {Zahid}, {Guillochon},
  {Mandel}, {Auchettl}, \& {Zabludoff}}]{2018ApJ...853...39G}
{Graur}, O., {French}, K.~D., {Zahid}, H.~J., {et~al.} 2018, \apj, 853, 39,
  \dodoi{10.3847/1538-4357/aaa3fd}

\bibitem[{{Guillochon} \& {Ramirez-Ruiz}(2013)}]{2013ApJ...767...25G}
{Guillochon}, J., \& {Ramirez-Ruiz}, E. 2013, \apj, 767, 25,
  \dodoi{10.1088/0004-637X/767/1/25}

\bibitem[{{Hills}(1975)}]{1975Natur.254..295H}
{Hills}, J.~G. 1975, \nat, 254, 295, \dodoi{10.1038/254295a0}

\bibitem[{{Kesden}(2012)}]{2012PhRvD..85b4037K}
{Kesden}, M. 2012, \prd, 85, 024037, \dodoi{10.1103/PhysRevD.85.024037}

\bibitem[{{Kochanek}(2016)}]{2016MNRAS.458..127K}
{Kochanek}, C.~S. 2016, \mnras, 458, 127, \dodoi{10.1093/mnras/stw267}

\bibitem[{{Komossa}(2015)}]{2015JHEAp...7..148K}
{Komossa}, S. 2015, Journal of High Energy Astrophysics, 7, 148,
  \dodoi{10.1016/j.jheap.2015.04.006}

\bibitem[{{Law-Smith} {et~al.}(2017{\natexlab{a}}){Law-Smith}, {MacLeod},
  {Guillochon}, {Macias}, \& {Ramirez-Ruiz}}]{2017ApJ...841..132L}
{Law-Smith}, J., {MacLeod}, M., {Guillochon}, J., {Macias}, P., \&
  {Ramirez-Ruiz}, E. 2017{\natexlab{a}}, \apj, 841, 132,
  \dodoi{10.3847/1538-4357/aa6ffb}

\bibitem[{{Law-Smith} {et~al.}(2017{\natexlab{b}}){Law-Smith}, {Ramirez-Ruiz},
  {Ellison}, \& {Foley}}]{2017ApJ...850...22L}
{Law-Smith}, J., {Ramirez-Ruiz}, E., {Ellison}, S.~L., \& {Foley}, R.~J.
  2017{\natexlab{b}}, \apj, 850, 22, \dodoi{10.3847/1538-4357/aa94c7}

\bibitem[{{Leloudas} {et~al.}(2019){Leloudas}, {Dai}, {Arcavi}, {Vreeswijk},
  {Mockler}, {Roy}, {Malesani}, {Schulze}, {Wevers}, \&
  {Fraser}}]{2019arXiv190303120L}
{Leloudas}, G., {Dai}, L., {Arcavi}, I., {et~al.} 2019, arXiv e-prints,
  arXiv:1903.03120.
\newblock \doarXiv{1903.03120}

\bibitem[{{Lodato} {et~al.}(2009){Lodato}, {King}, \&
  {Pringle}}]{2009MNRAS.392..332L}
{Lodato}, G., {King}, A.~R., \& {Pringle}, J.~E. 2009, \mnras, 392, 332,
  \dodoi{10.1111/j.1365-2966.2008.14049.x}

\bibitem[{{Mockler} {et~al.}(2019){Mockler}, {Guillochon}, \&
  {Ramirez-Ruiz}}]{2019ApJ...872..151M}
{Mockler}, B., {Guillochon}, J., \& {Ramirez-Ruiz}, E. 2019, \apj, 872, 151,
  \dodoi{10.3847/1538-4357/ab010f}

\bibitem[{{Moore} \& {Garaud}(2016)}]{2016ApJ...817...54M}
{Moore}, K., \& {Garaud}, P. 2016, \apj, 817, 54,
  \dodoi{10.3847/0004-637X/817/1/54}

\bibitem[{{Paxton} {et~al.}(2011){Paxton}, {Bildsten}, {Dotter}, {Herwig},
  {Lesaffre}, \& {Timmes}}]{2011ApJS..192....3P}
{Paxton}, B., {Bildsten}, L., {Dotter}, A., {et~al.} 2011, \apjs, 192, 3,
  \dodoi{10.1088/0067-0049/192/1/3}

\bibitem[{{Ramirez-Ruiz} \& {Rosswog}(2009)}]{2009ApJ...697L..77R}
{Ramirez-Ruiz}, E., \& {Rosswog}, S. 2009, \apjl, 697, L77,
  \dodoi{10.1088/0004-637X/697/2/L77}

\bibitem[{{Rees}(1988)}]{1988Natur.333..523R}
{Rees}, M.~J. 1988, \nat, 333, 523, \dodoi{10.1038/333523a0}

\bibitem[{{Steinberg} {et~al.}(2019){Steinberg}, {Coughlin}, {Stone}, \&
  {Metzger}}]{2019MNRAS.485L.146S}
{Steinberg}, E., {Coughlin}, E.~R., {Stone}, N.~C., \& {Metzger}, B.~D. 2019,
  \mnras, 485, L146, \dodoi{10.1093/mnrasl/slz048}

\bibitem[{{Stone} {et~al.}(2019){Stone}, {Kesden}, {Cheng}, \& {van
  Velzen}}]{2019GReGr..51...30S}
{Stone}, N.~C., {Kesden}, M., {Cheng}, R.~M., \& {van Velzen}, S. 2019, General
  Relativity and Gravitation, 51, 30, \dodoi{10.1007/s10714-019-2510-9}

\bibitem[{{Tejeda} {et~al.}(2017){Tejeda}, {Gafton}, {Rosswog}, \&
  {Miller}}]{2017MNRAS.469.4483T}
{Tejeda}, E., {Gafton}, E., {Rosswog}, S., \& {Miller}, J.~C. 2017, \mnras,
  469, 4483, \dodoi{10.1093/mnras/stx1089}

\bibitem[{{Timmes} \& {Swesty}(2000)}]{2000ApJS..126..501T}
{Timmes}, F.~X., \& {Swesty}, F.~D. 2000, \apjs, 126, 501,
  \dodoi{10.1086/313304}

\end{thebibliography}
\bibliographystyle{aasjournal}

\end{document}